\documentclass[%
 reprint,
 amsmath,amssymb,
 aps,
]{revtex4-1} 

\usepackage{graphicx}
\usepackage{dcolumn}
\usepackage{bm}

\usepackage{graphicx}
\usepackage{dcolumn}
\usepackage{bm}
\usepackage{epsfig}
\usepackage{color}
\usepackage{longtable}

\def\mem#1#2#3{  \left\langle #1 \left\vert  #2 \right\vert #3 \right\rangle   }

\newcolumntype{w}[1]{D{.}{.}{#1}}

\begin{document}

\preprint{}
%
%
%
%
\title{Atomic physics studies at the Gamma Factory at CERN}
%
%
%
%

\author{Dmitry Budker}
\affiliation{Helmholtz Institut, Johannes Gutenberg-Universit{\"a}t Mainz, 55128 Mainz, Germany}
\affiliation{Department of Physics, University of California, Berkeley, California 94720, USA}

\author{Jos\'e~R.~{Crespo L\'opez-Urrutia}}
\affiliation{Max-Planck-Institut f\"ur Kernphysik, D--69117 Heidelberg, Germany}

\author{Andrei Derevianko}
\affiliation{Department of Physics, University of Nevada, Reno, 89557, USA}

\author{Victor V.\,Flambaum} 
\affiliation{School of Physics, University of New South Wales,  Sydney 2052,  Australia}
\affiliation{Helmholtz Institut, Johannes Gutenberg-Universit{\"a}t Mainz, 55128 Mainz, Germany}
\affiliation{The New Zealand Institute for Advanced Study, Massey University Auckland, 0632 Auckland, New Zealand}

\author{Mieczyslaw Witold Krasny}
\affiliation{LPNHE, Sorbonne Universit\'{e}, Paris Diderot Sorbonne Paris Cit\'{e}, CNRS/IN2P3, Paris; France}
\affiliation{CERN, Geneva, Switzerland}

\author{Alexey Petrenko}
\affiliation{Budker Institute of Nuclear Physics, Novosibirsk, Russia}
\affiliation{CERN, Geneva, Switzerland}

\author{Szymon Pustelny}
\affiliation{M.~Smoluchowski Institute of Physics, Jagiellonian University, 30--348 Krakow, Poland}

\author{Andrey Surzhykov}
\affiliation{Physikalisch--Technische Bundesanstalt, D--38116 Braunschweig, Germany}
\affiliation{Technische Universit\"at Braunschweig, D--38106 Braunschweig, Germany}

\author{Vladimir~A.~Yerokhin}
\affiliation{Physikalisch--Technische Bundesanstalt, D--38116 Braunschweig, Germany}
\affiliation{Center for Advanced Studies, Peter the Great St. Petersburg Polytechnic University, 195251 St. Petersburg, Russia}

\author{Max Zolotorev}
\affiliation{E.\,O\,.~Lawrence Berkeley National Laboratory, Berkeley, CA 94720, USA}

\date{\today \\[0.3cm]}

%
%
%
%

\begin{abstract}
The Gamma Factory initiative proposes to develop novel research tools at CERN by producing, accelerating and storing highly relativistic, partially stripped ion beams in the SPS and LHC storage rings. By exciting the electronic degrees of freedom of the stored ions with lasers, high-energy narrow-band photon beams will be produced by properly collimating the secondary radiation that is peaked in the direction of ions' propagation. Their intensities, up to $10^{17}$ photons per second, will be several orders of magnitude higher than those of the presently operating light sources in the particularly interesting $\gamma$--ray energy domain reaching up to 400 MeV. This article reviews opportunities that may be afforded by utilizing the primary beams for spectroscopy of partially stripped ions circulating in the storage ring, as well as the atomic-physics opportunities afforded by the use of the secondary high-energy photon beams. The Gamma Factory will enable ground breaking experiments in spectroscopy and novel ways of testing fundamental symmetries of nature.  \end{abstract}
%
\maketitle

%
%
\section{Introduction}

The Gamma Factory (GF) is an ambitious proposal, currently explored within the CERN Physics Beyond Colliders program \cite{Krasny2015}. The proposal aims at developing a source of narrow-band photons with energies up to $\approx 400\,$MeV, with photon fluxes up to $\approx 10^{17}$ photons per second, exceeding those of the currently available $\gamma$--ray sources (Table \ref{tab:gammay_ray_sources}) by many orders of magnitude. 

In this paper, we briefly survey some of the new opportunities that may be afforded by the GF in atomic physics and related fields. 

\begin{table*}[t]
    \centering
    \begin{tabular}{l      c      c      c      c}
    \hline 
    \hline
    Facility name     & ROKK-1M    & GRAAL     & LEPS      & HI$\gamma$S \\
    \hline \\[-0.2cm]
     Location & Novosibirsk & Grenoble  & Harima & Duke \\
     Storage ring & VEPP-4M & ESRF & SPring--8 & Duke--SR \\
     Laser--photon energy (eV) & 1.17–-4.68 & 2.41-–3.53 & 2.41–-4.68 & 1.17–-6.53\\
     $\gamma$--beam energy (MeV)  & 100–-1600 & 550–-1500 & 1500–-2400 & 1–-100 (158) \\
     $\Delta E / E$  & 0.01 – 0.03 & 0.011 & 0.0125 & 0.008 -- 0.1\\
     Max on--target flux ($\gamma$/s)  & 10$^6$ & 3 $\times$ 10$^6$ & 5 $\times$ 10$^6$ & 10$^4$ -- 5 $\times$ 10$^8$\\
    \hline
    \hline
    \end{tabular}
    \caption{Parameters of existing $\gamma$--ray sources around the world, from Ref.~\cite{WELLER2009257}. All the listed sources are based on inverse Compton scattering from beams of electrons circulating in storage rings.}
    \label{tab:gammay_ray_sources}
\end{table*}

The GF is based on circulating partially stripped ions (PSI), i.e., nuclei with a few bound electrons rather than bare nuclei, in a high-energy storage ring. The electrons intrinsic to the PSI open new experimental possibilities
for physics studies as well as for ion-beam control and cooling. Successful injection and storage of relativistic PSI was demonstrated  at SPS and LHC \footnote{The names of these storage rings come from Super Proton Synchrotron and Large Hadron Collider. SPS is used as an injector for LHC.}, with decay times of $\approx$40\,hrs in the latter case \cite{Schaumann2019,AbramovIPAC2019}.
We note that the atomic-, plasma-  and astrophysics communities typically refer to PSI as highly charged ions (HCI)\cite{gillaspy_highly_2001,lopez-urrutia_emission_2014,beiersdorfer_highly_2015}; in the following we use both terms, PSI and HCI, interchangeably. 

The presence of bound electrons makes electronic transitions possible. For few-electron heavy ions, these are in general in the x-ray region, although fine and hyperfine interactions can also induce smaller splittings. 

The main idea of the GF is to send light from a laser beam head-on to a PSI beam with a high relativistic factor $\gamma$. In the ion frame, the energy of the incident photons is boosted by a factor of $2\gamma$, enabling spectroscopy of the ions with the use of the primary-photon beams. The PSI excited with the primary beam emit secondary photons, which, upon transformation to the laboratory frame, are predominantly emitted in the direction of propagation of the PSI. Their energy in the laboratory frame is boosted by another factor of $2\gamma$, and can be tuned by changing $\gamma$ and the energy of the of the laser photons. Tunable, high-energy secondary photon beams from the GF can be used in a variety of experiments.

In order to take full advantage of the GF as a novel research tool and avoid missing some of the unprecedented opportunities it will afford, it is important to survey possible uses and address the existing and future challenges that may arise in various fields. In this spirit, we present here, without any claim for completeness, some of the ideas, with full understanding that the realization of each one of them is a complex technical challenge.

For decades, lasers and traps for atoms and ions have been among the most useful tools in atomic physics. The GF is both: a light source, akin to a laser, delivering nearly monochromatic high-energy photons (collimated secondary beams), and also a giant ion trap, where the PSI are interrogated with the primary laser photons. The latter scenario is conceptually analogous to laser spectroscopy with ``normal'' ion traps \cite{Itano1987}, and has been realized in ion storage rings at low values of $\gamma$, for example, at the Test Storage Ring (TSR) \cite{Wolf1991} 
of the Max-Planck Institute for Nuclear Physics in Heidelberg and at the Experimental Storage Ring (ESR) of GSI Darmstadt \cite{Novotny2009}.  

\begin{figure}[b]\centering
	\includegraphics[width=0.9\linewidth]{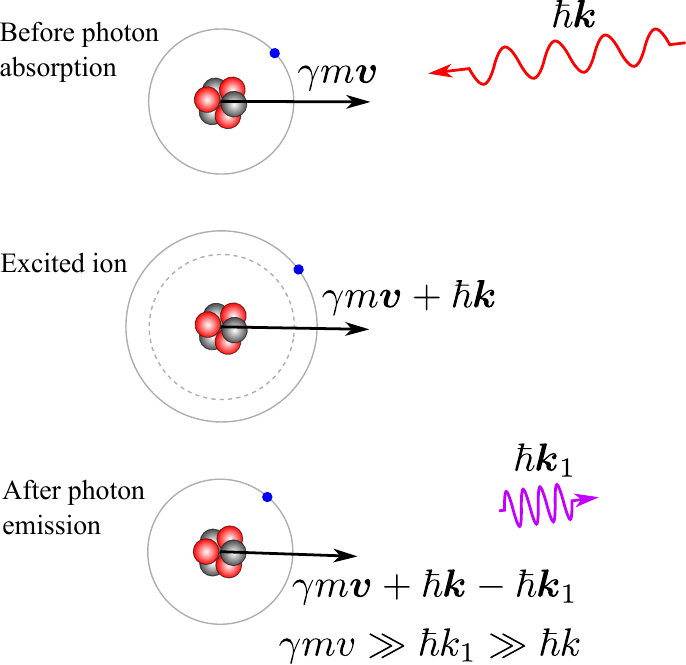}
	\caption{Photon scattering by a relativistic, partially stripped ion as observed in the laboratory frame of reference. Ion and photon momenta are indicated.}
    \label{fig:lab_frame}
\end{figure}

Some of the low-$\gamma$ experiments that are proposed for GF at CERN are also well suited for implementation at the future GSI/FAIR facility and are already in its research program. Techniques necessary for the GF, e.g., laser cooling, have been developed in TSR and ESR experiments \cite{Winters_2015}. This complementarity of GF and other facilities offers opportunities for fruitful collaboration.
%
%
\section{Basic principles of the Gamma Factory}
\label{sec:basic_principles}
\subsection{Primary beams}

The basic setup consists of optical photons (angular frequency $\omega$) that are directed head-on  or with a small angle onto a beam of ultra-relativistic ions (Fig.~\ref{fig:lab_frame}). In the ion frame of reference, the photon frequency is boosted to
\begin{equation}
    \label{eq:first_frequancy_transformation}
    \omega_0 = (1+\beta)\gamma\omega \approx 2\gamma\omega\,,
\end{equation}
where $\beta=v/c\approx 1$, and $v$ is the ion speed in the laboratory frame, related to the relativistic factor $\gamma$ according to
\begin{equation}
    \label{eq:gamma_factor}
    \gamma = \frac{1}{\sqrt{1-\beta^2}} = \frac{1}{\sqrt{1-(v/c)^2}}\,,
\end{equation}
and where $c$ is the speed of light.

\begin{figure}[t]\centering
	\includegraphics[width=0.9\linewidth]{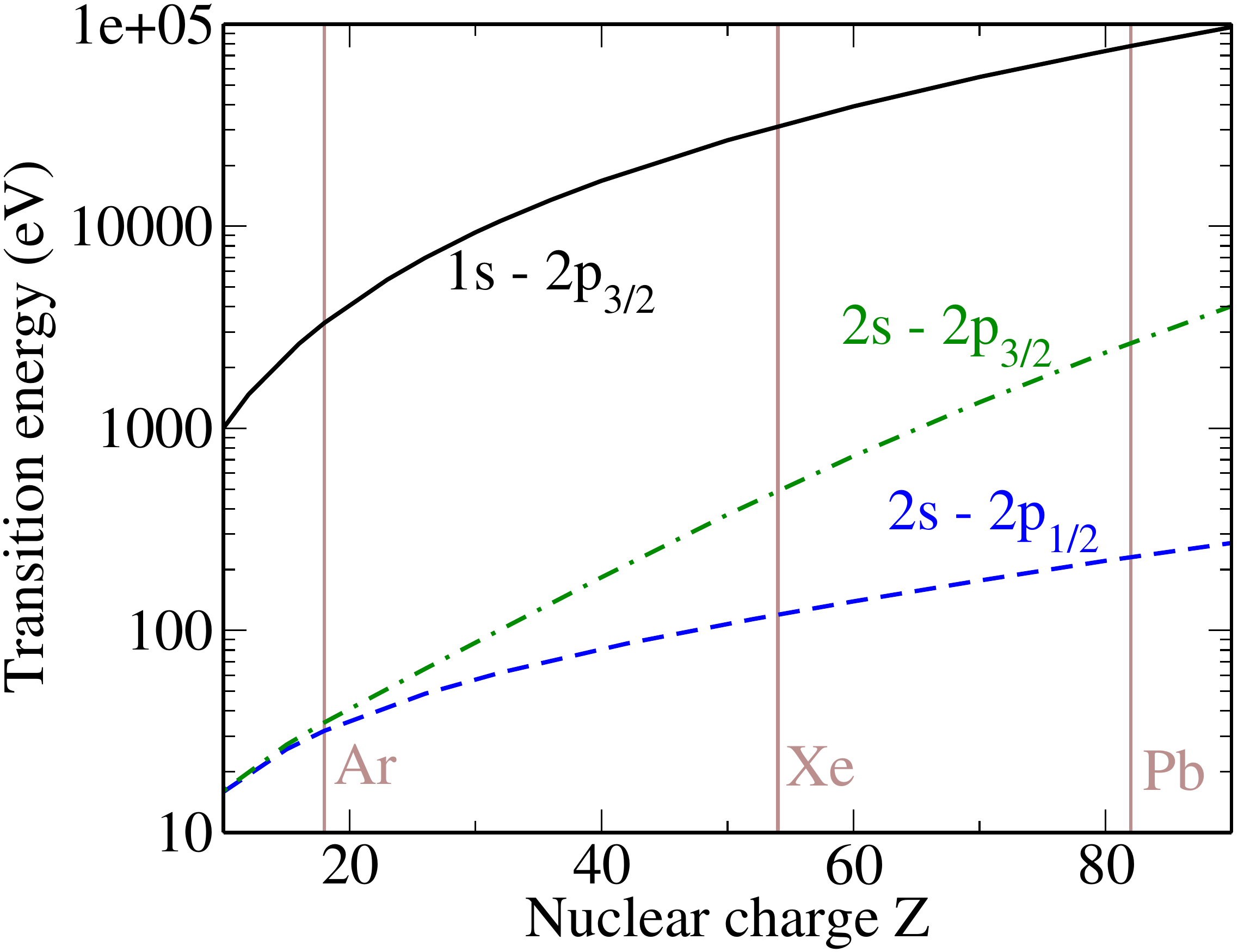}
	\caption{Energies of the $1s \to 2p_{3/2}$ transition in hydrogen--like ions (black solid line) as well as $2s \to 2p_{1/2}$ (blue dashed line) and $2s \to 2p_{3/2}$ (green dash--and--dotted line) transitions in lithium--like ions. The vertical lines mark hydrogen-- as well as lithium--like argon, xenon and lead ions.}
    \label{fig:GF_energies}
\end{figure}

Already in this first step, relativistic effects are extreme; with $\gamma\approx 3000$ available at the LHC and photon energies of up to $\approx 10$\,eV available with lasers, energies up to 60\,keV in the ion frame become available, allowing laser spectroscopy of and laser cooling (reducing momentum dispersion) on a wide range of electronic transitions in PSI that have not yet been studied, such as the $1s-2p$ transitions for hydrogen--like with $36<Z<80$ (Fig.\,\ref{fig:GF_energies}), where $Z$ is the atomic number.

\subsection{Secondary photons}
Staying for now with the example of the  $1s-2p$ transitions in hydrogenic ions, ions excited to the $2p$ state re-emit photons with isotropic angular distribution in the ion frame (when summed over all polarization directions; see, for example, Prob.~3.8. in Ref.~\cite{BudkerBookAtomPhysProblems}). 
Going back to the laboratory frame, the relativistic transformation has two important consequences (Fig.~\ref{fig:lab_frame}):
\begin{itemize}
    \item photon emission is concentrated in a small angle $\approx 1/\gamma$ in the direction of the ions' propagation; see, for example, Sec.\,14.3 in \cite{Jackson};
    \item the frequency (energy) of the photons re-emitted along the ions' propagation direction
    is boosted by another factor of $\approx 2\gamma$ \footnote{Including the angular dependence, the expression reads: $\omega'' = \gamma \omega' (1 + \beta\cos\theta')$, where $\theta'$ is is photon-emission angle in the ion frame of reference.}:
    \begin{equation}
    \omega^{\prime\prime} \approx 2\gamma\omega^{\prime} \approx 4\gamma^2\omega.
\end{equation}
\end{itemize}
The boost is smaller for photons emitted at an angle to the ion-momentum vector.
For heavy hydrogenic ions such as Pb$^{81+}$, laboratory-frame, secondary-photon energies up to $\approx 400$\,MeV can be achieved. This is the key idea of the Gamma Factory. An instructive analogy is that of undulator radiation, which is produced by relativistic electrons passing through cm-scale, alternating periodic magnetic structure. The static periodic field, whose period is contracted in the electron frame by the $\gamma$ factor, is seen as electromagnetic radiation by the electrons. In the GF case, the role of the undulator is played by the exciting laser light (and the electrons are bound rather than free). 

Two key points of the GF scheme have to be mentioned. First, resonant electronic excitations in PSI have orders of magnitude larger photon-scattering cross sections than bare ions or electrons, ensuring large secondary-photon fluxes. The resonant nature of the laser-ion interaction enables controlling the ionic internal states, and subsequently laser cooling of the ions in the storage ring. Second, in the GF scheme, tuning the secondary-photon energy is possible by combining tuning of the relativistic factor $\gamma$ and the choice of the electronic transitions excited in the PSI by the up-boosted optical laser.  
Some of the anticipated parameters of the GF are listed in Table \ref{tab:GF_parameters}.

\begin{table}[t]
    \centering
    \begin{tabular}{c c}
    \hline 
    \hline
    Parameter & Value \\
    \hline \\[-0.2cm]
    Ion $\gamma$ factor     &  10 -- 2900 \\[0.1cm]
    Ion species             & Pb$^{q+}$ as an example \\[0.1cm]
    Transverse beam radius  & 16 $\mu$m \\[0.1cm]
    Number of ions in a bunch & 10$^{8}$ \\[0.1cm]
    Number of bunches in the ring & 1232 \\[0.1cm]
    Effective repetition rate         & 10 MHz \\[0.1cm] 
    Ion energy spread       & 10$^{-4}$ \\[0.1cm] 
    RMS bunch length        & 7.9 cm  \\[0.1cm]
    Normalized emittance    & 1.6 $\mu$m \\ [0.1cm] 
    Circumference of LHC    & 26.7 km \\
    \hline
    \hline
    \end{tabular}
    \caption{Representative parameters of the Gamma Factory at CERN. $q$ denotes the charge state of the ions. The numbers are presented for Pb ions.}
    \label{tab:GF_parameters}
\end{table}
%
%

%
%
\section{Partially stripped ions: State of the art}
\label{sec:PSI_general}

Partially stripped (or highly charged) ions offer unique ways for exploring various fundamental questions in modern science. In the realm of atomic physics, these ions serve as natural laboratories to probe few--electron systems exposed to strong electromagnetic fields produced by nuclei. For instance, an electron in the $1s$ ground state of hydrogen--like uranium U$^{91+}$ experiences an electric field strength of about 10$^{16}\,$V/cm; this value far exceeds those attainable by focusing short--pulsed laser beams, and approaches the so--called Schwinger critical field 
\begin{equation}
  E_s=m^2c^3/(e\hbar) \approx 1.3\times 10^{16}\  \textrm{V/cm},
\end{equation}
at which electron--positron pairs can be spontaneously created. In the presence of such strong fields, the energies of the atomic states in PSI differ from those of neutral atoms, see Table\,\ref{tab:my_label}. The electrons are tightly bound, with ionization energies that scale as the square of the ion charge, $Z^2$, and that can reach 100\,keV for heavy systems like U$^{91+}$. A similar $Z^2$ scaling holds for energies of transitions between electronic states with different principal quantum numbers $n$. For PSI, these transition energies can be in the x-- and even $\gamma$--ray spectral regions unlike transitions in neutral atoms that lie in the visible and ultraviolet (UV) domains.

Many interactions that are usually suppressed in atoms become remarkably strong in PSI. The couplings of electron spin and orbital motion as well as with nuclear moments lead to fine-- (fs) and hyperfine--structure (hfs) splittings which scale as $\Delta E_{fs} \propto Z^4$  and $\Delta E_{hfs} \propto Z^3$. For medium-- and high--$Z$ ions, these splittings are in the range of $\Delta E_{fs} \approx$~keV and $\Delta E_{hfs} \approx$~eV. Also quantum--electrodynamics (QED) effects yield large energy shifts $\Delta E_{QED}\propto Z^4$ or having even steeper $Z$ dependence, reaching values of several hundred eV for the heaviest systems \cite{GuS05}. 

\begin{table}[t]
    \centering
    \begin{tabular}{l l}
    \hline
    \hline 
    Transition energy $\Delta E_{n n'}$  &  $\propto (Z\alpha)^2$ \\[0.2cm]
    Fine--structure splitting            &  $\propto (Z\alpha)^4$ \\[0.2cm]
    Hyperfine--structure splitting       &  $\propto \alpha(Z\alpha)^3m_e/m_p$ \\[0.2cm]
    Lamb shift                           &  $\propto \alpha(Z\alpha)^4$ \\
    \hline
    \hline
    \end{tabular}
    \caption{$Z$--scaling of atomic characteristics for hydrogen--like ions. Each of the energies is a product of the scaling factors given in the table, a numerical factor, and $m_ec^2$, where $m_e$ is the electron mass and $c$ is the speed of light. $m_p$ is the proton mass.  }
    \label{tab:my_label}
\end{table}

The above discussion shows that probing the level structure of heavy PSI tests atomic systems in the critical nonperturbative QED regime. However, lying in the x-- and $\gamma$--ray domain, bound--state transitions in PSI cannot be reached with conventional lasers. In the majority of modern experiments, therefore, the excited ionic states are produced in various collisional processes and their subsequent radiative decay is observed with solid--state detectors \cite{EiS07,GuF11}. Such experiments, performed usually at electron-beam ion traps (EBIT) as well as ion storage rings, have certain accuracy restrictions. For example, the uncertainty of the so far most accurate measurement of the $1s$ Lamb shift in hydrogen--like uranium, $\Delta E_{1s-QED} = 460.2$~eV, is 4.6 eV \cite{GuS05}, and its further reduction remains a challenging task.

Spectroscopy of PSI in the high--$Z$ region attracted continued theoretical and experimental attention during the last three decades. This interest was triggered to a large extent by the  experiment \cite{schweppe:91} on Li--like uranium at the Lawrence Berkeley Laboratory's Bevalac accelerator. In this experiment, uranium U$^{89+}$ ions were produced from a 95\,MeV beam of U ions by beam--foil stripping. The ions were then magnetically separated and transported to a second foil where the $2s_{1/2} \to 2p_{1/2}$ transition with energy 280\,eV  was excited. Argon--gas cells installed at appropriate viewing angles filtered the emitted photons 
using the $L_{23}$ absorption edge. In order to determine their energy, the fraction transmitted to a set of detectors behind the gas cells was registered as a function of the ion-energy-dependent Doppler shift under a certain viewing angle. The sub--eV accuracy achieved in this experiment clearly demonstrated the need for the QED calculations of second order in the fine--structure constant $\alpha$. Moreover, these calculations had to be performed without any expansion in the nuclear binding--strength parameter $Z\alpha$, since the value of this parameter approaches unity for high--$Z$ ions, and in particular for Pb and U. This was a challenge to the theory, which required developments of new calculational methods and which was finally met only several decades later. The calculations were performed by several authors, notably, by the G\"oteburg \cite{lindgren:95:pra,persson:96:2el}, Notre--Dame \cite{blundell:93:b,sapirstein:01:lamb,SaC11}, and St.~Petersburg \cite{yerokhin:00:prl,YeI06,ArS07} groups. The main motivation for those studies was testing the bound--state QED theory in a regime of strong electron--nucleus Coulomb interactions. Even now, after nearly three decades since the experimental achievement \cite{schweppe:91}, this interesting regime is not yet accessible by any other means \cite{Beiersdorfer2010,Volotka2013}. An exhaustive recent review of this field can be found in Ref.\,\cite{Indelicato2019}. 

\begin{table}
    \centering
\begin{tabular}{lcw{10.08}ll}
    \hline
    \hline
  \multicolumn{1}{l}{Ion}
      & \multicolumn{1}{c}{Transition}
                 & \multicolumn{1}{c}{Energy [eV]}
                        & \multicolumn{2}{c}{Reference}\\
\hline\\[-5pt]
$^{208}$Pb$^{81+}$& $2p_{3/2}$ -- $1s$ &    77\,934.59\,(26)  & theo & \cite{yerokhin:15:Hlike} \\[0.2cm]
$^{238}$U$^{91+}$ & $2p_{3/2}$ -- $1s$ &   102\,173.1\,(4.3)  & exp & \cite{Gumberidze2003}\\
                  &                    &   102\,175.10\,(53)  & theo & \cite{yerokhin:15:Hlike} \\[0.2cm]
$^{238}$U$^{90+}$ & $1s2p\ ^1P_1$ -- $1s^2\, ^1S_0$ & 100\,626.0 \,(35) & exp & \cite{briand:90} \\

&                                      &   100\,610.89\,(65)   & theo & \cite{artemyev:05:pra} \\
&                                      &   100\,610.68\,(54)   & theo & \cite{kozhedub:19}\\[0.2cm]
%
%
    \hline
    \hline
\end{tabular}
\caption{Experimental (exp) and theoretical (theo) $2p-1s$ transition energies in heavy hydrogen-- and helium--like ions.}
\label{tab:2p1s}
\end{table}
\begin{table*}[t]
    \centering
\begin{tabular}{lcw{8.11}ll}
    \hline
    \hline
  \multicolumn{1}{l}{Ion}
      & \multicolumn{1}{c}{Transition}
                 & \multicolumn{1}{c}{Energy [eV]}
                        & \multicolumn{2}{c}{Reference}\\
\hline\\[-5pt]
    Pb$^{79+}$ & $1s^22p_{1/2}$ -- $1s^22s$   & 230.823\,(47)(4)  & theo & \cite{YeS18,YeI06,Kozhedub2010} \\
               &                              & 230.76\,(4)        & theo & \cite{SaC11} \\[0.2cm]
    Bi$^{80+}$ & $1s^22p_{1/2}$ -- $1s^22s$   & 235.809\,(53)(9)   & theo & \cite{YeS18,YeI06,Kozhedub2010} \\
               &                              & 235.72\,(5)        & theo & \cite{SaC11} \\[0.2cm]
    U$^{90+}$ & $1s2p\ ^3P_0$ -- $1s2s\ ^3S_1$&  260.0\,(7.9)      & exp & \cite{munger:86}\\
              &                               &  252.01\,(27)      & theo & \cite{artemyev:05:pra} \\
              &                               &  251.94\,(11)      & theo & \cite{kozhedub:19}\\[0.2cm]
    U$^{89+}$ & $1s^22p_{1/2}$ -- $1s^22s$    & 280.645\,(15)      & exp & \cite{BeC05} \\
              &                               & 280.775\,(97)(28)  & theo & \cite{YeS18,YeI06,Kozhedub2010} \\[0.2cm]
    \hline
    \hline
    \end{tabular}
\caption{Experimental (exp) and theoretical (theo) $2p_{1/2}-2s$ transition energies in
heavy ions. If the energy is given with two uncertainties, the first one is the estimate of the theoretical error, whereas the second one is due to the error of the nuclear charge root-mean-square (rms) radius.}
    \label{tab:2p12s}
\end{table*}

The measurement reported in Ref.\,\cite{schweppe:91} was surpassed in accuracy by later experiments on various PSI transitions (see, for example, Refs.\,\cite{Indelicato2019, Beiersdorfer2014PRL} and references therein). These experiments and dedicated theoretical investigations enabled the presently most stringent tests of the bound--state QED in the strong--field regime. However, a persistent obstacle for these tests are the strong nuclear--size contributions to the binding energy. These corrections cannot be accurately predicted in the absence of detailed knowledge of nuclear parameters such as the charge root-mean-square (RMS) radius and nuclear magnetization distribution. Similar to the proton--size puzzle presented by apparently contradictory results of high--resolution laser spectroscopy in hydrogen atoms and in muonic hydrogen \cite{2010Pohl,2013Antognini,2013AntogniniAnnals}, our fragmentary knowledge of the nuclear structure hindered high--precision QED tests in the high--$Z$, high--field, nonperturbative regime. This was realized for HCI already in the late 1990's after measurements of the hyperfine structure of hydrogen--like ions that showed serious inconsistencies with predictions \cite{Klaft1994PRL,crespo_lopez-urrutia_direct_1996,Seelig1998PRL,Crespo1998PRA,Beiersdorfer2001PRA}. A practical solution to this conundrum, the method of \emph{specific differences}, was developed by Shabaev and co-workers \cite{Shabaev2001PRL}. It is based on the smooth structure of the electronic wave function in the neighborhood of the nucleus, as well as on the detailed analysis of the $n$--scaling of the $1s$ electron density as a function of the principal quantum number $n$. By measuring, for example, the Lamb shifts in transitions from a certain $ns$ level, the nuclear overlap can be extrapolated to other $n's$ orbitals. This allows largely removing uncertainties due to nuclear--size contributions to the binding energy. Further improvements of the method \cite{Volotka2012PRL,Ginges2018PRA} will enable better tests of the QED.

We discuss how experiments with the GF can contribute to advancing the HCI spectroscopy in the following Sec.\,\ref{subsec:Spectroscopy of PSI}.

\section{Physics cases for the Gamma Factory}
\label{sec:physics_cases}
\subsection{Spectroscopy of partially stripped ions}
\label{subsec:Spectroscopy of PSI}

The GF will open up intriguing opportunities for the spectroscopy of high--$Z$ PSI. The unique feature of this facility is that the Doppler tuning of the photon energy enables \textit{direct} access to the bound--state ionic transitions in the x--ray domain. To emphasize this feature, let us first briefly summarize what is presently known about the spectra of high--$Z$ partially stripped ions. In the high--$Z$ regime, the electron-electron interaction (correlation) is a small contribution to the total binding energy, and its relative size is suppressed by a factor of $1/Z$, so the transition energy is basically determined by the differences of one--electron energies. For that reason, similar transitions in isonuclear ions in different charge states are close in energy (for example, $2p \to 1s$ decay of an H--like ion lies close to $1s 2p \to 1s^2$ of the corresponding He--like ion, and this again is close to the Li-like $1s2s2p \to 1s^22s$ transition). Such line series are often called \emph{satellite spectra}.

The most intense and well-resolved lines in the x--ray spectra are due to the Lyman--$\alpha$, $2p\to 1s$ transitions. Their energies for high-$Z$ ions are of the order of 100~keV (Table~\ref{tab:2p1s}) and for that reason, they are difficult to measure precisely. The best accuracy achieved for the transition in U$^{91+}$ is currently 5~eV~\cite{GuS05}. As mentioned in Sec.\,\ref{sec:PSI_general}, further improvement of accuracy remains a challenging experimental problem. The corresponding accuracy of theoretical predictions for these transitions is in a sub--eV range, being an order of magnitude better than the experimental precision.

The situation is rather different for the $2p\to 2s$ and $2p_{3/2}\to 2p_{1/2}$ transitions. Their energies are lower than those of Lyman--$\alpha$ lines, for the heaviest HCI being $\sim 3$-$4$~keV for the $2p_{3/2}\to \{2s,2p_{1/2}\}$ and $\sim 0.2$-$0.3$~keV for the $2p_{1/2}\to 2s$ decays. It makes these \textit{intrashell} transitions accessible for experimental determination using x--ray crystal spectrometers at a sub--eV accuracy level. Accurate experimental values for these transitions in various HCI are shown in Tables~\ref{tab:2p12s}, \ref{tab:2p32s}, and \ref{tab:2p32p1} summarizing the $2p_{1/2}\to 2s$, $2p_{3/2}\to 2s$, and $2p_{3/2}\to 2p_{1/2}$ transitions, respectively. The best accuracy of 0.015~eV was attained for the $2p_{1/2}\to 2s$ transition energy in Li--like uranium \cite{BeC05} using crystal spectrometry at the Lawrence Livermore National Laboratory's EBIT. Comparison with theory probed the two--loop QED effects and provided  currently the best test of QED theory in the strong nuclear Coulomb field \cite{YeI06}.

\begin{table*}[t]
    \centering
\begin{tabular}{lcw{8.12}ll}
    \hline
    \hline
  \multicolumn{1}{l}{Ion}
      & \multicolumn{1}{c}{Transition}
                 & \multicolumn{1}{c}{Energy [eV]}
                        & \multicolumn{2}{c}{Reference}\\
\hline\\[-5pt]
    Pb$^{79+}$ & $1s^22p_{3/2}$ -- $1s^22s$   & 2\,642.26\,(10)      & exp & \cite{ZhN08}\\
               &                              & 2\,642.220\,(46)(4)  & the & \cite{YeS18,YeI06,Kozhedub2010} \\[0.2cm]
    Bi$^{80+}$ & $1s^22p_{3/2}$ -- $1s^22s$   & 2\,788.139\,(39)     & exp & \cite{BeO98}  \\
               &                              & 2\,788.127\,(52)(10) & the & \cite{YeS18,YeI06,Kozhedub2010} \\[0.2cm]
    Th$^{87+}$ & $1s^22p_{3/2}$ -- $1s^22s$   &  4\,025.23\,(15)     & exp & \cite{beiersdorfer:95}  \\
               &                              &  4\,025.41\,(10)(10) & the & \cite{YeS18,YeI06,Kozhedub2010} \\
               &                              &  4\,025.25\,(7)      & the & \cite{SaC11} \\[0.2cm]
    Th$^{86+}$ & $1s^22s2p_{3/2}\ P_1$ -- $1s^22s^2\ ^1S_0$
                                              &  4\,068.47\,(13)     & exp & \cite{beiersdorfer:95}  \\[0.2cm]
    U$^{90+}$ & $1s2p\ ^3P_2$ -- $1s2s\ ^3S_1$&  4\,509.71\,(99)     & exp & \cite{trassinelli:09}\\
              &                               &  4\,510.03\,(26)     & the & \cite{artemyev:05:pra} \\
              &                               &  4\,509.88\,(11)     & the & \cite{kozhedub:19}\\[0.2cm]
    U$^{89+}$ & $1s^22p_{3/2}$ -- $1s^22s$    &  4\,459.37\,(25)(10)  & exp & \cite{BeK93}   \\
              &                               &  4\,459.580\,(94)(31) & the & \cite{YeS18,YeI06,Kozhedub2010} \\[0.2cm]
    U$^{88+}$ & $1s^22s2p_{3/2}\ P_1$ -- $1s^22s^2\ ^1S_0$
                                              &  4\,501.72\,(21)     & exp & \cite{BeK93}\\
    \hline
    \hline
    \end{tabular}
\caption{Experimental (exp) and theoretical (the) $2p_{3/2}\to 2s$ transition energies in
heavy ions.}
    \label{tab:2p32s}
\end{table*}

In most experiments performed so far, the electronic states of PSI were excited through collisional processes, and their subsequent radiative decays observed. Such experiments were carried out at accelerators/storage rings (for example, \cite{schweppe:91,gumberidze_quantum_2005}) and EBITs (for example, \cite{beiersdorfer_highly_2015}) equipped with high--resolution spectrometers. Another approach used in recent experiments \cite{NaT13} is resonant coherent excitation (RCE) of relativistic uranium ions channeled through the periodic field of an oriented crystal. More recent experiments use x-ray free-electron lasers (XFEL) \cite{epp_soft_2007,bernitt_unexpectedly_2012} and monochromatic synchrotron radiation \cite{rudolph_x-ray_2013} to resonantly excite transitions up to over 13\,keV photon energy \cite{Epp2015}.  

The GF offers a unique alternative to EBIT, RCE, XFEL, and synchrotron experiments. In the GF, the transitions of interest will be \textit{directly} driven by the (Doppler--boosted) primary laser beam. More specifically, photoexcitation of the ground state into the $1s^22p_{1/2}$ excited state of lithium--like Pb$^{79+}$ is proposed as the first proof--of--principle experiment at the GF \cite{Krasny2019PoP}. With the planned parameters of this experiment presented in Table~\ref{tab:PoP_parameters}, one can expect that the $2s \to 2p_{1/2}$ transition energy will be measured with a relative accuracy of about $10^{-4}$, which is better than the accuracy of the theoretical prediction, see Table~\ref{tab:2p12s}. To the best of our knowledge, this will become the first experimental observation of the $1s^22p_{1/2}$ -- $1s^22s$ transition in Pb$^{79+}$; its main purpose, however, will be to demonstrate the feasibility of the GF for precision x--ray spectroscopy of PSI. 

An important advantage of the GF is the ability to excite a wide selection of electronic transitions. In particular, one can envisage extensions of the proof--of--principle measurement towards transitions involving higher excited states of Pb$^{79+}$, $1s^22s \to 1s^2np_{j}$ with $n \ge 2$. There is little knowledge about these highly--excited states, and the GF experiments will provide valuable experimental data that can contribute to further investigations of isotope-sensitive nuclear--size and QED effects in few--electron systems. 

Besides the lithium--like ions, the Doppler--boosted primary photon beams at the GF can be employed to explore many other PSI. Of special interest is, for example, the electric--dipole--forbidden transition
$1s^2 \, 2s^2 \, 2p_{1/2} \, {}^2P_{1/2} \to 1s^2 2s^2 2p_{3/2} {}^2P_{3/2}$ in heavy boron--like systems. Similar to the lithium--like case, the splitting between $2 {}^2P_{1/2}$ and $2 {}^2P_{3/2}$ fine--structure levels is of purely relativistic nature. Therefore, the transition between these two levels provides a perfect testing ground for the relativistic and QED effects that are not masked
in this case by (often overwhelming) non--relativistic contributions. For high--$Z$ ions, for which these effects become most pronounced, the transition energy $\hbar \omega = E_{P_{3/2}} - E_{P_{1/2}}$ is a few keV. As mentioned above, such a transition can be easily accessed at the GF, thus opening a unique opportunity for testing higher--order QED corrections in strong electromagnetic fields.

Until now we have discussed how the (Doppler--boosted) primary photon beams can be used for the spectroscopy of partially stripped heavy ions. However, high--precision measurements of the transition energies in PCI can be based also on the analysis of re--emitted secondary photons.  That is, high--resolution $\gamma$ spectroscopy using flat--crystals spectrometers, whose principles were developed at the Institute Laue-Langevin \cite{Kessler2001,Rainville2005,Dewey2006}, could determine the energies of the doubly--boosted emitted MeV photons with accuracy better than 100 parts--per-billion (ppb). This is due to the fact that the lattice--spacing determination for appropriate Si crystals is better than 0.5 ppb \cite{Massa2009}. In combination with the much more accurate knowledge of the primary laser frequency, this will result in a determination of the electronic transition energy in the circulating PSI with a sub--parts--per--million accuracy, an improvement of two orders of magnitude relative to the current storage-ring measurements of high--$Z$ ions.

In order to illustrate the advantages of the PSI spectroscopy based on the measurements of the secondary photons let us briefly revisit formulas from Sec.~\ref{sec:basic_principles}. Given a bound--bound transition with energy $\hbar\omega_0$ in a PSI moving with the relativistic factor of Eq.\,\eqref{eq:gamma_factor}, a laser photon with energy $\hbar\omega$ will be absorbed and re-emitted if the resonance condition of Eq.\,\eqref{eq:first_frequancy_transformation} is met in the moving ion frame of reference. This can be used to \emph{select} a group near a particular energy (relativistic factor $\gamma$) within the PSI bunch out of the broader energy spread.

\begin{table}[t]
    \centering
\begin{tabular}{lcw{8.12}ll}
    \hline
    \hline
  \multicolumn{1}{l}{Ion}
      & \multicolumn{1}{c}{Transition}
                 & \multicolumn{1}{c}{Energy [eV]}
                        & \multicolumn{2}{c}{Reference}\\
\hline\\[-5pt]
    U$^{87+}$  & $1s^22s^22p\ ^2P_{3/2}$ -- $^2P_{1/2}$     & 4\,087.02\,(17)      & exp & \cite{beiersdorfer:98}\\
               &                                            & 4\,087.59\,(41)      & the & \cite{ArS13} \\[0.2cm]
    U$^{83+}$  & $1s^22s^22p^5\ ^2P_{3/2}$ -- $^2P_{1/2}$   & 3\,913.54\,(16)      & exp & \cite{beiersdorfer:98}\\
               &                                            & 3\,913.76\,(2)       & the & \cite{volotka:19} \\[0.2cm]
    \hline
    \hline
    \end{tabular}
\caption{Experimental (exp) and theoretical (the) $2p_{3/2} \to 2p_{1/2}$ fine-structure splitting energies in heavy ions.}
    \label{tab:2p32p1}
\end{table}

The subsequent radiative decay of PSI leads to the emission of the (secondary) photons whose frequency in the laboratory frame is, including the angular dependence,
\begin{equation}
    \omega''= \frac{4\gamma^2 \omega}{1+(\gamma\theta)^2} \, .
\end{equation}
Here we have assumed that the laboratory-frame photon-emission angle $\theta$, defined with respect to the PSI beam, is small, $\theta \approx 1/\gamma$. State--of--the-art flat-crystal transmission spectrometers have angular selectivity of the order of 50\,nrad FWHM \cite{Kessler2001} (the spectrometer of Ref.\,\cite{Kessler2001} operated at photon energies up to 6\,MeV). If we assume an uncertainty for $\Delta\theta\approx5\times10^{-8}$\,rad, such a spectrometer could select the photon energy re-emitted at a small angle $\theta<5\times10^{-8}$\,rad by PSI moving at $\gamma\approx3000$ with a relative uncertainty given by the following expression:
\begin{equation}
    \frac{\Delta\omega''}{\omega''} \approx 2\gamma^2\theta \Delta\theta \approx 4.5\times10^{-8} \, .
\end{equation}
If this value would constitute the largest contribution to the experimental error in the determination of the ion transition energy, it would mark a significant improvement over the current best electronic transition-energy determinations in high--$Z$ PSI. 

%
%
%
%
We note that by selecting the photons at larger angles $\theta$, it is, in principle, possible to perform metrology along similar lines but detecting, for example, ultraviolet or visible photons. A small fraction of photons ($\sim 1/\gamma$) will be emitted at large angles $\sim$~1~rad in the laboratory frame of reference. The frequency of these photons is comparable to the frequency of incoming laser light. For example, at 90~degree angle the frequency of emitted radiation will be twice the frequency of the laser light $\omega$ used to excite the ion.

If the PSI transition energy $\omega_{0}$ is known beforehand by some other means with an uncertainty $\Delta\omega_{0}/\omega_{0}$, the method delivers instead $\omega''$ with a similar relative error bar. Thus, the GF can be used either (i) to study electronic transitions in PSI with high accuracy if the re--emitted photons are measured with a commensurate relative error, or (ii) to generate tunable gamma rays of excellent energy definition, since we assume that the PSI kinetic energy and the laboratory laser-photon energy can be freely chosen.

%
%
\subsection{Atomic parity violation}

Atomic parity violation (APV) is a powerful probe of the electroweak sector of the Standard Model, as well as a tool to search for physics beyond the Standard Model. APV tests are unique in their sensitivity to neutral currents, and also complementary to collider experiments, since they probe the domain of low-momentum-transfer interactions. While most APV experiments have focused on neutral or singly ionized atoms, several APV measurements in PSI have been proposed.

At the microscopic level, APV is predominantly caused by the weak interaction between electrons and the quarks mediated by the $Z^0$ boson within the nucleus, and mixes electronic levels of opposite parity. In contrast to neutral systems, in PSI, the mixing coefficient
\begin{equation}
    \label{eq:PNC_mixing_coefficient}
    \eta = \frac{\mem{\Psi_s}{{\hat H}_{w}}{\Psi_p}}{E_p - E_s - i\Gamma/2}
\end{equation}
is strongly enhanced since the matrix element of the weak interaction Hamiltonian ${\hat H}_{w}$ scales as $Z^5$, as opposed to $Z^3$ for neutral atoms \footnote{Note that the requirement of the absence of CP violation renders the weak-interaction matrix element to be purely imaginary if one neglects the effect of the finite transition width $\Gamma$. A consequence of this is the absence of a permanent electric-dipole moment of the mixed states.}. Here the numerator represents the matrix element of the weak interaction between the $s$ and $p$ states, while the denominator is their complex energy gap with $\Gamma$ representing the width of the transition.  

\begin{table}[t]
    \centering
    \begin{tabular}{l l}
    \hline
    \hline 
    Parameter  \hspace{2cm} &  Value   \\[0.1cm]
    \hline\\[-0.3cm]
    Crossing angle                 & 2.6\,deg \\[0.1cm]
    Ion magnetic rigidity          & 787\,T$\cdot$m \\[0.1cm]
    Ion $\gamma$ factor            & 96.3   \\[0.1cm]
    Ion beam horizontal RMS size at IP & 1.3\,mm \\[0.1cm]
    Ion beam vertical RMS size at IP   & 0.8\,mm \\[0.1cm]
    Ion revolution frequency           & 43.4 kHz \\[0.1cm]
    Laser photon energy                & 1.2\,eV \\[0.1cm]
    Laser pulse repetition rate                    & 40\,MHz \\[0.1cm]
    Laser pulse energy                 & 5\,mJ \\
    \hline
    \hline
    \end{tabular}
    \caption{Planned experimental parameters of the proof--of--principle experiment  at the CERN SPS accelerator, aimed at the direct photoexcitation of the $2 \; {}^2S_{1/2} \, \to \, 2 \; {}^2P_{1/2}$ transition in lithium-like lead. IP denotes the interaction point.}
    \label{tab:PoP_parameters}
\end{table}

\begin{figure}[t]\centering
	\includegraphics[width=0.9\linewidth]{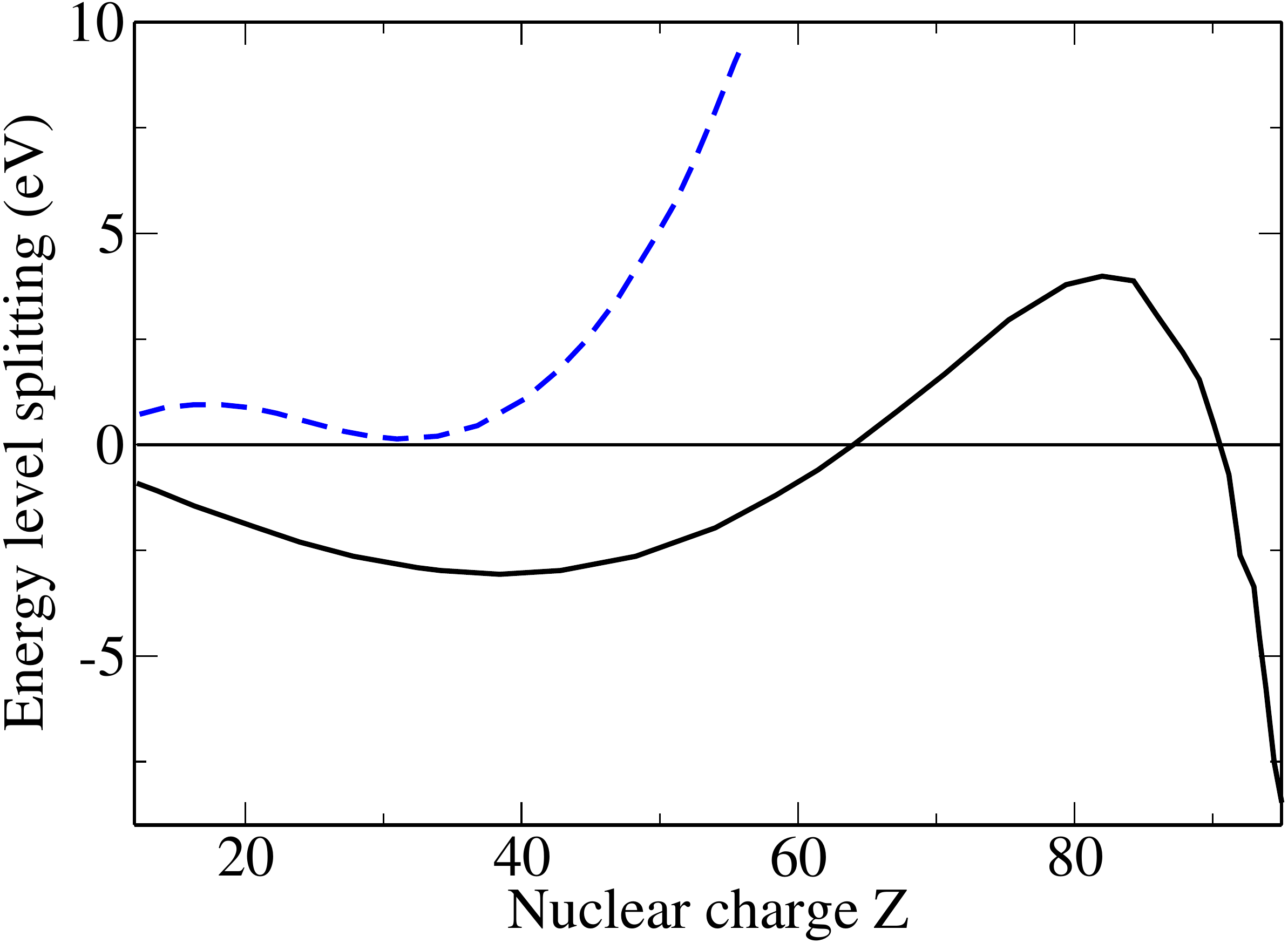}
	\caption{Energy splitting between $1s 2s \; ^1S_0$ and $1s 2s \; ^3P_0$ (black solid line) as well as between $1s 2s \; ^1S_0$ and $1s 2s \; ^3P_1$ (blue dashed line) levels of helium-like ions as a function of nuclear charge Z.}
    \label{fig:He-like-energies}
\end{figure}

Such a significant enhancement of the parity-violating (PV) mixing is caused by the larger electron-nucleus overlap in PSI, the fact that the weak charge of the nucleus scales as $Z$, and the scaling of the matrix element with the electron momentum $p \propto Z$ (see Ref.\,\cite{Khriplovich1991} for further details). Another advantage of PSI is that their energy spectra can be finely ``tuned'' by varying the nuclear charge  and their charge state (i.e., the number of electrons bound to the nucleus). In particular, one can observe effects of crossing of ionic energy levels. This crossing, which happens when two ionic states have almost the same energy, can be used to further enhance the PV mixing [Eq.\,\eqref{eq:PNC_mixing_coefficient}]. For example, the levels $1s2s\,^1S_0$ and $1s2p\,^3P_0$ of helium-like ions are known to be nearly degenerate for nuclear charges $Z = 64$ and $Z = 92$ (Fig.\,\ref{fig:He-like-energies}). The non-monotonic behavior of the energies of $2 {}^1S_0$ and $2 {}^3P_0$ states is caused by the interplay of the electron-electron interaction and relativistic and QED contributions, each with different $Z$--scaling, see Ref.\,\cite{FeS11} and references therein. The theoretical calculations of these corrections are still controversial and, hence, experimental measurements of the  $2 {}^1S_0$--$2 {}^3P_0$ energy splittings are highly desired. At the GF, this energy splitting can be determined from the combined measurements of the $2 {}^1S_0$--$2 {}^1P_1$ and $2 {}^3P_0$--$2 {}^1P_1$ transitions. Thus, spectroscopy of singly-excited states of helium-like ions can significantly contribute both to APV and to atomic-structure studies in the high-$Z$ domain. 

Over the past decade, several proposals were made to measure the PV mixing between ionic levels with the goal of accessing the weak interaction effects in these atomic systems. Most of the proposals, however, rely on measuring laser--induced transitions from excited ionic states. For example, the single--photon $1s2s {}^1S_0$--$1s2s {}^3S_1$ \cite{Shabaev2010} and two-photon $1s2s {}^1S_0$--$1s2p {}^3P_0$ \cite{Surzhykov2011} transitions in helium-like ions are currently discussed as possible candidates for APV experiments at storage rings. A serious drawback of these and similar proposals is the short lifetime of excited states of PSI, which usually does not exceed 10$^{-12}$~s. A promising alternative to these approaches is excitation of an ion from its ground state. For example, we propose to directly drive the transition between the levels $1s^2 \; {}^1S_0,\;F=I$ and $1s2s \; {}^1S_0,\;F = I$ of helium-like ions with nonzero nuclear spin $I \ne 0$. The $1 \; ^1S_0$--$2 \; {}^1S_0$ transition may proceed either via the parity--conserving (hyperfine--induced) magnetic dipole M1 channel or via the parity--violating electric dipole E1 excitation. The latter becomes possible due to the PV mixing between the hyperfine sublevels of the $1s2s \; {}^1S_0$ and $1s2p \; 2{}^3P_1$ states. This mixing is mainly induced by the weak interactions within nuclei, whose dominant part \cite{Flambaum1980} comes from the so--called anapole moment \cite{Zeldovich1958}. In order to measure the $2 \; ^1S_0$--$ {}^3P_1$ PV mixing and hence to study properties of the nuclear anapole moment, one could observe the circular dichroism in the $1\;^1S_0$--$ 2 {}^1S_0$ transition, i.e., the difference in excitation rates for the right-- and left--circularly polarized light. Because of the (relatively) large parity mixing coefficient $|\eta| \sim 10^{-10}$ and the existence of a stable isotope ${}^{77}_{34}$Se, the Se$^{32+}$ ion is one of the most promising candidates for the this photoexcitation experiment \cite{FeS11}.  The $1 {}^1S_0$--$2 {}^1S_0$ transition energy for this ion is 11.6 keV which can be easily accessed at the GF.

A similar approach to APV measurements was proposed for hydrogen-like ions \cite{Zolotorev1997}. For these ions, the weak interaction between electron and nucleus leads to the mixing of the $2s$ and $2p_{1/2}$ ionic states whose energies differ just by the Lamb shift. This system is particularly attractive due to the large parity mixing and tractable electronic structure theory. To observe this mixing one needs to drive the $1s \to 2s$ (Stark-induced  E1 + PV E1) transition and observe the circular dichroism. Since the transition energy increases as $Z^2$ with the nuclear charge, it reaches the hard x-ray domain already for medium-$Z$ ions. GF would critically enable such experiments.

%
%
\subsection{Extracting neutron skin from the measurement of parity violation in iso--nuclear sequence of highly--charged ions}

Neutron skin or halo refers to the difference in neutron and proton distributions inside the nucleus. While charge (proton) distributions are well measured in Coulomb-scattering experiments  and measurements of isotope shifts in electronic and muonic atoms, neutron distributions are not. Neutron distributions can be extracted from the measurements probing the weak force as the (nuclear-spin-independent part of the) weak interaction predominantly couples atomic electrons and neutrons. 

Conversely, better knowledge of neutron skins should enable more precise APV experiments because the neutron-skin effect ultimately limits the extraction of new physics from APV \cite{Derevianko2001,Brown2009,Derevianko2002}. Knowing the neutron skin in Pb fixes nuclear-model parameters \cite{Brown2009} and thus this will have an important impact on constraining neutron-skin uncertainties in interpreting experiments with neutral atoms such as Cs, Yb, Dy, Rb and Fr \cite{Cadeddu2019,Viatkina2019}, singly ionized ions, and diatomic molecules where APV measurements are planned or ongoing. 

Thus, the choice of ${}^{208}$Pb for APV experiments at the LHC could be especially advantageous. There are already direct measurements of ${}^{208}$Pb neutron skin in nuclear physics experiments (see, for example, Refs.\,\cite{Tarbert2014,Abrahamyan2012}). In those works, the difference in RMS radii between the proton and neutron distributions $\Delta R_{np}$ was extracted with $\approx$30\% accuracy. 
We will explore the possibility of extracting the neutron skin from APV measurements with a higher accuracy with various ions of Pb.  For example, a comparison of APV in H-like, He-like, Li-like Pb (and potentially Pb ions with more electrons) will provide for independent measurements of the neutron-skin effect, as the correction in individual ions is proportional to the same $\Delta R_{np}$. Importantly, at the GF, being an accelerator facility, the investigations of the neutron skins in unstable nuclei will be also feasible. With this, one can significantly extend the range of the isotopes where the neutron-skin effects can be studied. For example, this could include the low-lying isomeric state $^{229m}$Th that is currently in the focus of fundamental-physics investigations \cite{Thirolf2019}.

It is worth noting that improving the knowledge of neutron skins has a direct bearing on the understanding of the neutron-matter equation of state \cite{Piekarewicz2019,Brown2000} and will improve the interpretation of binary neutron-star mergers \cite{Fattoyev2018}, such as GW170817 detected by LIGO/Virgo \cite{Abbott2017} via gravitational waves and in follow-up multi-messenger observations in a broad spectral range of  electromagnetic radiation. 


%
\subsection{Laser polarization of partially stripped ions}

During the laser-light interaction with the relativistic PSI, the ions will cycle between the ground and excited states, in the course of which, light polarization can be transferred to the PSI electrons and also to the nucleus when the latter has a nonzero spin. Such processes are referred to as \emph{optical pumping}. Spin-polarized PSI open intriguing possibilities for both atomic and nuclear experiments that crucially rely on the polarization degrees of freedom. 

Optical pumping can be accomplished on a single path of a PSI bunch through the interaction region with the laser light; however, how one can utilize the PSI polarization depends on whether it will survive a round trip in the storage ring.

It is still an open question of accelerator dynamics whether polarization can survive the machine's turning magnets. Assuming that it does, this opens possibilities for fundamental physics experiments with such PSI, potentially measuring the ionic and nuclear electric dipole moments (EDM) that violate both parity and time-reversal invariance. Searches for both static \cite{Pretz2013} and oscillating \cite{Pretz2019} EDM are of great current interest, see, for example, \cite{Safronova2018}.

However, even if the polarization cannot survive a round trip in the ring, optical pumping still offers exciting possibilities of fixed-target experiments with polarized PSI.

Leaving aside the technical challenges of practical realization, one can also consider colliding-beam experiments with polarized PSI. In this case, the two counter-propagating PSI beams would both need to be polarized in the straight section of the accelerator containing the interaction region (collision point). One can speculate that collisions of polarized heavy nuclei might open novel inroads to the study of quark-gluon plasma \cite{Jacak2012}.

Returning to APV, optical pumping would enable the study of nuclear spin-dependent APV effect, which allows access to parity-violating nuclear anapole moments and measurements of the weak meson coupling constants (see Ref.\,\cite{Safronova2018} and references therein).  

Optical polarization of the PSI in relativistic storage ring was considered in Ref.\,\cite{Prozorov2003}. Optical pumping of H-like ions with circularly polarized light was proposed to produce the polarization of both ionic nuclei and electronic shells \cite{Bondarevskaya2011}. While the first optical-pumping experiments were performed for low-$Z$ systems, no ion polarization has as yet been achieved in the high-$Z$ domain. The schemes for production and implementation of spin-polarized ions at the GF are currently under discussion. 
%
%
\subsection{Interaction of vortex light with ion beams}

During the last decade, light beams with a helical phase front that carry orbital angular momentum (OAM) came into focus of intense theoretical and experimental attention. While such ``twisted beams'' are routinely produced today across the terahertz, infrared, visible and UV ranges \cite{Knyazev2018}, the generation of OAM $\gamma$--rays is still an open task. Recently, Compton back--scattering was proposed as a process which produces twisted x--rays \cite{Jentschura2011}. Resonant scattering of (initially twisted) optical photons by PSI is a potential scenario for production of twisted $\gamma$--rays.  Resonant scattering experiments at the GF will help to investigate the feasibility of this approach. 

Apart from the investigation of the feasibility of the production of twisted $\gamma$-rays, there are scientific opportunities arising from the interaction of the primary twisted light beams with the PSI. 
In contrast to conventional plane waves, twisted light allows one to modify the relative strength of transition multipoles by ``switching off,'' for example, the leading dipole terms. The theory of the excitation of electric, magnetic, and mixed-moment transitions using vortex light beams was recently developed \cite{Solyanik2019,Schulz2019}. The extension of the light sources to ultra-short wavelength ranges will enable studies of the deeply bound electronic states and also of the nuclear degrees of freedom. Moreover, significant suppression of the AC Stark shift induced by twisted light, as demonstrated in the infrared with a single-ion model system \cite{Schmiegelow2016}, would make OAM x- and $\gamma$--rays a valuable tool for high-precision spectroscopy of HCI and nuclei.

%
%
\subsection{PSI spectroscopy in strong external fields}

The region where PSI interact with the laser beam can be placed into an external field, for instance, a tunable strong transverse magnetic field. In the frame of the PSI, this field appears as orthogonal transverse electric and magnetic fields enhanced by the relativistic factor $\gamma$. With modern high--field magnets \cite{Battesti2018}, electric fields in the PSI frame of 10$^{12}$\,V/cm or even higher are conceivable, allowing studying electric polarizabilities of the PSI and manipulating their energy levels via the Stark effect. The ability to apply external fields is also important for the fundamental--symmetry tests discussed above. 

\subsection{Tests of special relativity}

Precision laser spectroscopy of ions in a storage ring opens a possibility of testing special relativity, for example, time dilation, as was successfully done with mildly relativistic PSI at GSI and Heidelberg, see Refs.\,\cite{Novotny2009,Botermann2014} and references therein. At the GF, these tests can potentially be extended into the ultrarelativistic regime, which would require development of methods for high-precision determination of the ions' relativistic factors (see Sec.\,\ref{subsec:Spectroscopy of PSI}). Improved test for a possible anisotropy of the one-way maximum attainable speed \cite{Bergan2020} may also be possible.

\subsection{Photon scattering by highly charged ions}

So far we have discussed mostly the use of the primary photon beams at the GF. Even more atomic physics studies can be carried out with the secondary beams. For example, the x- and $\gamma$--ray photons, produced at the GF, can be used not only for the bound-state spectroscopy of HCI but also for scattering studies. Of special interest here is the elastic scattering of high-energy photons by ions. This scattering may proceed via three main channels: nuclear Thomson scattering, Rayleigh scattering by an electron cloud and Delbr\"uck scattering by quantum vacuum. In the keV range, the Rayleigh and Delbr\"uck contributions are the dominant ones and, hence, the analysis of the angular and polarization properties of scattered photons can provide valuable information about the structure of the HCI and coupling to the quantum vacuum. Moreover, the theoretical analysis of both, Rayleigh and especially Delbr\"uck channels still remains an open challenge. These GF studies may complement recent elastic scattering experiments performed at the PETRAIII synchrotron facility \cite{Blumenhagen2016}.

\section{Outlook}
\label{sec:outlook}
In July of 2018, the Gamma Factory at CERN made a major step from idea to reality with beams of H-like and He-like lead having been circulated in the SPS for several minutes. The H-like beam was further injected into the LHC, where it circulated for hours, from which the beam lifetime of over 40 hrs was inferred \cite{Schaumann2019}. The next crucial step is the proof-of-principle experiment \cite{Krasny2019PoP} that should validate the entire GF concept. 

Of course, there are many challenges that would need to be addressed before the GF is able to realize its full potential. These include, for example, realization of laser cooling of the PSI in the ultrarelativistic regime, precise matching of the ion and photon energy spread to achieve efficient PSI excitation, and development of techniques for precision calibration of the PSI energy. In conjunction with the latter, we note that some of the most precise techniques for energy calibration of ultrarelativistic particles (with relative uncertainty of a few parts in $10^5$) are, indeed, based on the determination of the secondary-photon energy in the inverse Compton-scattering geometry similar to that of the GF \cite{Klein2002,Abakumova2014,Abakumova2013}.

In this paper, we sketched of some of the possible novel opportunities that will open when the Gamma Factory is realized at CERN, starting with the proof-of-principle experiment planned to be carried out within months of this writing \cite{Krasny2019PoP}. While it is essentially certain that the opportunities we have covered are only a fraction of what can be done with this fundamentally new tool, we hope this provides a starting point for further exploration. 

From the time of Galileo, new instruments have allowed us to expand our horizons and make amazing discoveries about Nature and the Universe. We believe the Gamma Factory is poised to become such a paradigm-shifting tool. 

\section{Acknowledgements}
\label{sec:acknowledgements}

The authors are grateful for helpful discussions with and suggestions by Julian Berengut, Savas Dimopoulos, Alejandro Garcia, Felix Karbstein, Olga Kocharovskaya, Nancy Paul, J\"org Pretz, Fritz Riehle, Ferdinand Schmidt-Kaler, Yannis Semertzidis, Valeriy~G.~Serbo,  Vladimir Shabaev, Thomas St\"ohlker, and Bogdan Wojtsekhowski. This work was supported in part by the DFG Project ID 390831469:  EXC 2118 (PRISMA+ Cluster of Excellence), by the US National Science Foundation (award No.\,1912465), and by the Foundation for the Advancement of Theoretical Physics and Mathematics ``BASIS''.

%
%
%
%
%

%
%
%
%
\bibliography{./GF_AP_references}

\end{document}